\documentclass[11pt, a4paper]{article}

\usepackage[T1]{fontenc}
\usepackage[utf8]{inputenc} 
\usepackage{graphicx,color} 
\usepackage{epsfig}
\usepackage{times} 
\usepackage{amsmath}
\usepackage{amssymb}
\usepackage{caption}
\usepackage{subcaption}
\usepackage{stackrel}

\usepackage{url}
\usepackage{multirow}
\usepackage{cite}

\usepackage{algorithm}
\usepackage{pgf}
\usepackage{pgfplots}
\usepgfplotslibrary{fillbetween}
\usepackage{tikz}
\usetikzlibrary{arrows,automata}
\usetikzlibrary{intersections}
\usetikzlibrary{calc}
\usetikzlibrary{shapes}
\usetikzlibrary{positioning}
\usetikzlibrary{decorations.text}
\pgfdeclarelayer{background}
\pgfdeclarelayer{foreground}
\pgfsetlayers{background,main,foreground}
\usetikzlibrary{calc,patterns,decorations.pathmorphing,decorations.markings}

\definecolor{mycol}{RGB}{19,48,128}
\usepackage{hyperref} 
\hypersetup{
	colorlinks = true,
	citecolor = mycol,
	linkcolor = mycol,,
	urlcolor = black
}

\usepackage{enumitem}
\renewcommand{\theenumi}{\arabic{enumi}}

\renewcommand{\theenumii}{\arabic{enumii}}

\newcommand{\NN}{{\mathcal N}}

\newtheorem{lemma}{Lemma}

\newtheorem{corollary}{Corollary}

\newcommand{\rr}{\mathbb{R}}

\newcommand{\eqdef}{\triangleq}

\newcommand{\proof}{\noindent{\em Proof}.~}
\newcommand{\QED}{$\Box$}

\newcommand{\beqar}{\begin{eqnarray}}
\newcommand{\eeqar}{\end{eqnarray}}
\newcommand{\beqarno}{\begin{eqnarray*}}
\newcommand{\eeqarno}{\end{eqnarray*}}
\newcommand{\ba}[1]{\begin{array}{#1}}
\newcommand{\ea}{\end{array}}

\newcommand{\col}{\mathop{\rm col}\nolimits}

\newcommand{\smallmat}[1]{\left[ \begin{smallmatrix}#1 \end{smallmatrix} \right]}

\newcommand{\tx}{{\theta_x}}
\newcommand{\ty}{{\theta_y}}

\begin{document}

	\title{\LARGE An L-BFGS-B approach for linear and nonlinear system identification under $\ell_1$ and group-Lasso regularization}
	
	\author{Alberto Bemporad 
    \thanks{The author is with the IMT School for Advanced Studies, Piazza San Francesco 19, Lucca, Italy. Email: \texttt{\scriptsize alberto.bemporad@imtlucca.it}.
    This work has received support from the European Research Council (ERC), Advanced Research Grant COMPACT (Grant Agreement No. 101141351)
}}

\maketitle
\thispagestyle{empty}
	
\begin{abstract}
In this paper, we propose a very efficient numerical method based on the L-BFGS-B algorithm for identifying linear and nonlinear discrete-time state-space models, possibly under $\ell_1$- and group-Lasso regularization
for reducing model complexity. 
For the identification of linear models, we show that, compared to classical linear subspace methods, 
the approach often provides better results, is much more general in terms of the loss and regularization terms used (such as penalties for enforcing system stability), and is also more stable from a numerical point of view. The proposed method not only enriches the existing set of linear system identification tools but can also be applied to identifying a very broad class of parametric nonlinear state-space models, including recurrent neural networks. We illustrate the approach on synthetic and experimental datasets and apply it to solve a challenging industrial robot benchmark for nonlinear multi-input/multi-output system identification. A Python implementation of the proposed identification method is available in the package \texttt{jax-sysid}, available at \url{https://github.com/bemporad/jax-sysid}.
\end{abstract}


\section{Introduction}
Model-based control design requires, as the name implies, a dynamical model of the controlled process, typically a linear discrete-time one. Learning dynamical models that explain the relation between input excitations and the corresponding observed output response of a physical system over time, a.k.a. ``system identification'' (SYSID), has been investigated since the 1950s~\cite{SL19}, 
mostly for linear systems~\cite{Lju99}. In particular, subspace identification methods like N4SID~\cite{VD94} and related methods~\cite{VD95} have been used with success in practice and are available in state-of-the-art software tools like the System Identification Toolbox for MATLAB~\cite{Lju01} and the Python package SIPPY~\cite{AVDBP18}. The massive progress in supervised learning methods for regression, particularly feedforward and recurrent neural networks (RNNs), has recently boosted research 
in \emph{nonlinear} SYSID~\cite{LATS20,PAGLRS23}. In particular, RNNs are black-box models that can capture the system's dynamics concisely in a deterministic nonlinear state-space form, and techniques have been proposed for identifying them~\cite{SSBCV17}, including those based on autoencoders~\cite{MB21}. 

In general, control-oriented modeling requires finding a balance between model simplicity and representativeness, as the complexity of the model-based controller ultimately depends on the complexity of the model used, in particular in model predictive control (MPC) design~\cite{BBM17,MRD18}. Therefore, two conflicting objectives must be carefully balanced: maximize the quality of fit and minimize the complexity of the model (e.g., reduce the number of states and/or sparsify the matrices defining the model). For this reason, when learning small parametric nonlinear models for control purposes,
quasi-Newton methods~\cite{LN89}, due to their convergence to very high-quality solutions, 
and extended Kalman filters (EKFs)~\cite{PF94,Bem23b} can be preferable to simpler (stochastic) gradient descent methods like Adam~\cite{KB14} and similar variants.
Moreover, to induce sparsity patterns, $\ell_1$-regularization is often employed, although it leads to non-smooth objectives; optimization methods were specifically introduced to handle $\ell_1$-penalties~\cite{AG07} and, more generally, non-smooth terms~\cite{SKS12,STP17,Bem23e,ABO23}. 

In this paper, we propose a novel approach to solve linear and nonlinear SYSID problems, possibly under $\ell_1$ and group-Lasso regularization, based on the classical L-BFGS-B algorithm~\cite{BLNZ95}, an extension of the L-BFGS algorithm to handle bound constraints, whose software implementations are widely available. To handle such penalties, we consider the positive and negative parts of the parameters defining the model and provide two simple technical results that allow us to solve a regularized problem with well-defined gradients. To minimize the open-loop simulation error, we eliminate the hidden states by creating a condensed-form 
of the loss function, which is optimized with respect to the model parameters and initial states.
By relying on the very efficient automatic differentiation capabilities of the JAX library~\cite{JAX} to compute gradients, we will first show that the approach can be effectively used to identify linear state-space models on both synthetic and experimental data, analyzing the effect of group-Lasso regularization for reducing the number of states or for selecting the most relevant input signals. 
We will also show that the approach is much more stable from a numerical viewpoint and often provides better results than classical linear subspace methods like N4SID, which, in addition,
cannot handle non-smooth regularization terms, non-quadratic losses, and constraints on model parameters. 
We will also show how to identify open-loop stable linear models via additional non-quadratic regularization terms.
Finally, we will apply the proposed method in a nonlinear SYSID setting to solve the challenging industrial robot benchmark for black-box nonlinear multi-input/multi-output SYSID proposed in~\cite{WGUR22}, by identifying a recurrent neural network (RNN) model in residual form under $\ell_1$-regularization
largely extending the preliminary results presented in~\cite{Bem23d}.

\section{System identification problem}
Given a sequence of input/output training data $(u_0,y_0)$, $\ldots$, $(u_{N-1},y_{N-1})$,
$u_k\in\rr^{n_u}$, $y_k\in\rr^{n_y}$, we want to identify a deterministic state-space model in 
the following form
\begin{equation}
    \begin{split}
        x_{k+1}&=Ax_k+Bu_k+f_x(x_k,u_k;\tx)\\
        \hat y_k&=Cx_k+Du_k+f_y(x_k,u_k;\ty)
    \end{split}
    \label{eq:model}
\end{equation}
where $k$ denotes the sample instant, $x_k\in\rr^{n_x}$ is the vector of hidden states,
$A,B,C,D$ are matrices of appropriate dimensions,
$f_x:\rr^{n_x}\times\rr^{n_u}\to\rr^{n_x}$, and 
$f_y:\rr^{n_x}\times\rr^{n_u}\to\rr^{n_y}$ are nonlinear
functions parametrized by $\tx\in\rr^{n_{\theta x}}$ and $\ty\in\rr^{n_{\theta y}}$, respectively. 

The training problem we want to solve is (cf.~\cite{Bem23e}):
\begin{subequations}
\begin{equation}
    \min_{z,x_1,\ldots,x_{N-1}}r(z)+\displaystyle{\frac{1}{N}\sum_{k=0}^{N-1} \ell(y_{k},Cx_k+Du_k+f_y(x_k,u_k;\ty))}\nonumber\\[-.6em]\label{eq:training-cost-reg}\\
\end{equation}
\begin{equation}
\begin{aligned}
    \mbox{s.t.}\quad & x_{k+1}=Ax_k+Bu_k+f_x(x_k,u_k;\tx)\label{eq:dyn-constraints-reg}\\
    &k=0,\ldots,N-2\nonumber\nonumber\\
    & z\eqdef [x_0'\ \col{A}'\ \col{B}'\ \col{C}'\ \col{D}'\ \tx'\ \ty']'\nonumber
\end{aligned}
\end{equation}
\label{eq:training}%
\end{subequations}
where the optimization vector $z\in\rr^{n_z}$ collects the entries of the initial state $x_0$ and of the model parameters $A,B,C,D,\tx,\ty$, with dimension $n_z=n_x+(n_x+n_u)(n_x+n_y)+n_{\tx}+n_{\ty}$, 
$\col$ is the vectorization operator stacking matrix columns,
$\ell:\rr^{n_y}\times\rr^{n_y}\to\rr$ is a differentiable loss function, and
$r:\rr^{n_z}\to\rr$ is a regularization term. 

After eliminating $x_k$, $k=1,\ldots,N-1$,
by replacing $x_{k+1}$ as in~\eqref{eq:dyn-constraints-reg}, Problem~\eqref{eq:training} 
can be rewritten as the unconstrained (and in general nonconvex) nonlinear programming (NLP) problem 
\begin{subequations}
\begin{eqnarray}
    &&\min_{z} f(z)+r(z)
    \label{eq:cost-reg}\\
    &&\hspace*{-2em}f(z)=\frac{1}{N}\sum_{k=0}^{N-1} \ell(y_{k},Cx_k+Du_k+f_y(x_k,u_k;\ty)).    
    \label{eq:training-cost-condensed}
\end{eqnarray}
\label{eq:training-condensed}%
\end{subequations}

\subsection{Special cases}
\subsubsection{Linear state-space models}
\label{sec:lin-sysid}
A special case of~\eqref{eq:training} is system identification of linear
state-space models based on the minimization of the simulation error
\begin{subequations}
\begin{eqnarray}
    \min_{z}&&r(z)+\displaystyle{\frac{1}{N}\sum_{k=0}^{N-1} \|y_{k}-Cx_k-Du_k\|_2^2}\label{eq:lin-sysid-cost}\\
    \mbox{s.t.}&& x_{k+1}=Ax_k+Bu_k,\ k=0,\ldots,N-2\label{eq:lin-sysid-dyn}\\
    && z\eqdef [x_0'\ \col{A}'\ \col{B}'\ \col{C}'\ \col{D}']'.\nonumber
\end{eqnarray}
\label{eq:lin-sysid}%
\end{subequations}
Specific model structures, such as $y_k=[I\ 0]x_k$ or other sparsity patterns, can be imposed by simply removing components of $z$. Moreover, bound constraints on model parameters can be enforced in~\eqref{eq:lin-sysid}; for example, \emph{positive linear systems} can be identified by constraining $A,B,C,D$ (and, possibly, $x_0$) to be nonnegative.

\subsubsection{Training recurrent neural networks}
\emph{Recurrent neural networks} (RNNs) are special cases of~\eqref{eq:model} in which $f_x,f_y$ are multi-layer feedforward neural networks (FNNs) parameterized by weight/bias terms $\tx\in\rr^{n_{\theta x}}$ and $\ty\in\rr^{n_{\theta y}}$, respectively,
and $A=0$, $B=0$, $C=0$, $D=0$. In particular, $f_x$ is a FNN with linear output function and  $L_x-1$ layers parameterized by weight/bias terms $\{A_i^x,b_i^x\}$
(whose components are collected in $\tx$), 
$i=1,\ldots,L_x$, and activation functions $f_i^x$, $i=1,\ldots,L_x-1$,
and similarly $f_y$ by weight/bias terms $\{A_i^y,b_i^y\}$, 
$i=1,\ldots,L_y$ (forming $\ty$) and activation functions $f_i^y$, $i=1,\ldots,L_y-1$,
followed by a possibly nonlinear output function $f_{L_y}^y$~\cite{Bem23b,Bem23e}.
We denote by $n_1^x,\ldots,n_{L_x-1}$ and
$n_1^y,\ldots,n_{L_y-1}$ the number of neurons in the hidden layers
of $f_x$ and $f_y$, respectively. 

\subsubsection{$\ell_2$ and $\ell_1$ regularization}
To prevent overfitting, both standard $\ell_2$ and $\ell_1$ regularization terms can be introduced.
In Section~\ref{sec:results} we will use the elastic-net regularization
\begin{equation}
    r(z)=\frac{1}{2}\left(\rho_\theta\left\|\Theta\right\|_2^2+\rho_{x}\|x_0\|_2^2\right)+\tau\left\|\Theta\right\|_1
\label{eq:l2-l1-regularization}
\end{equation}
where $\Theta$ collects all the model parameters
related to the state-update and output function, i.e., $z=[x_0'\ \Theta']'$, $\rho_\theta>0$, $\rho_x>0$, and $\tau\geq 0$. While $\ell_1$-regularization promotes model sparsity, 
it also makes the objective function in~\eqref{eq:training} non-smooth. The quadratic $\ell_2$-penalty term is also beneficial from an optimization perspective, in that it regularizes the Hessian matrix of the objective function. 

\subsubsection{Group-Lasso regularization}
To effectively reduce the order $n_x$ of the identified model~\eqref{eq:model},
the group-Lasso penalty~\cite{YL06} 
\begin{subequations}
\begin{equation}
r_g(z)=\tau_g\sum_{i=1}^{n_x}\|I_iz\|_2
\label{eq:group-Lasso-x}
\end{equation}
can be included in~\eqref{eq:cost-reg}, where $\tau_g\geq 0$ and $I_i$ is the submatrix formed by collecting the rows of the identity matrix of order $n_z$ corresponding to the entries of $z$ related to the initial state $x_{0i}$, the $i$th column and $i$th row of $A$, the $i$th row of $B$, and the $i$th column of $C$.
In the case of recurrent neural networks, the group also includes the $i$th columns of the weight matrices $A_1^x$ and $A_1^y$ of the first layer of the FNNs
$f_x$, $f_y$, the $i$th row of the weight matrix $A_{L_x}$, and $i$th entry of the bias term $b_{L_x}$ of the last layer of $f_x$, respectively~\cite{Bem23e}. 

Similarly, the group-Lasso penalty
\begin{equation}
r_g(z)=\tau_g\sum_{i=1}^{n_u}\|I_iz\|_2
\label{eq:group-Lasso-u}
\end{equation}%
\label{eq:group-Lasso}%
\end{subequations}
can be used to select the input channels that are most relevant to describe the dynamics of the system,
where now $I_i$ is the submatrix formed by collecting the rows of the identity matrix of order $n_z$ corresponding to the entries of $z$ related to the $i$th column of $B$ and $i$th column of $D$. 
This could be particularly useful, for example, when identifying Hammerstein models in which the input enters the model through a (possibly large) set of nonlinear basis functions.
In the case of recurrent neural networks, the $i$th group also contains the $(n_x+i)$th column of the weight matrices $A_1^x$ and $A_1^y$ of the first layer of the FNNs $f_x$, $f_y$. 
Clearly, group-Lasso penalties can be combined with $\ell_2$- and $\ell_1$-regularization.

We finally remark that \emph{rather arbitrary} (possibly nonconvex) loss functions $\ell$ can be 
handled by the approach proposed in this paper, as we use nonlinear programming to solve Problem~\eqref{eq:training}.

\subsection{Handling multiple experiments}
To handle multiple experiments $\{u^j_0,y^j_0,\ldots,u^j_{N_j-1},y^j_{N_j-1}\}$, $j=1,\ldots,M$,
Problem~\eqref{eq:training} can be reformulated as in
\cite[Eq.~(4)]{Bem23e} by optimizing with respect to $x_0^1,\ldots$, $x_0^M$, $A,B,C,D,\tx,\ty$,
or just $A,B,C,D,\tx,\ty$ with $x_0^j=0$, $\forall j=1,\ldots,M$. Alternatively, as suggested in~\cite{MB21,BTS21,Bem23b}, we
can introduce an initial-state encoder
\begin{equation}
    x_0^j=f_e(v_0^j;\theta_e) 
\label{eq:x0-encoder}
\end{equation}
where $v_0\in\rr^{n_v}$ is a measured vector available at time 0, such as a collection of 
$n_a$ past outputs, $n_b$ past inputs, and/or other measurements
that are known to influence the initial state of the system, 
and $f_{e}:\rr^{n_v}\times\rr^{n_{\theta_e}}\to\rr^{n_x}$ is
a FNN parameterized by a further optimization vector $\theta_e\in\rr^{n_{\theta_e}}$ to be learned jointly with $A,B,C,D,\tx,\ty$. 

\subsection{Stability constraints}
\label{sec:stability}
Although system instability is largely discouraged by minimizing the open-loop simulation error over the entire duration of the experiment(s), and the regularization terms also discourage the possible occurrence of unobservable modes, when learning linear models~\eqref{eq:lin-sysid} the identified matrix $A$ may have eigenvalues outside the unit circle, for example in the case of short experiments. Without loss of generality, as justified by next Lemma~\ref{lemma:stability}, asymptotic stability can be enforced by adding the following constraint
\begin{equation}
    \|A\|^2_2<1
\label{eq:stability-constraint}
\end{equation}
in~\eqref{eq:training-condensed}, where $\|A\|_2$ denotes the spectral norm of $A$.

\begin{lemma}
\label{lemma:stability}
Let $\Sigma\eqdef (A,B,C,D)$ be an asymptotically stable discrete-time linear system. Then there exists a transformation $T$ such that the equivalent system $\bar \Sigma\eqdef (\bar A,\bar B,\bar C,\bar D)$ is such that $\|\bar A\|_2<1$, where $\bar A=TAT^{-1}$, $\bar B=TB$, $\bar C=CT^{-1}$, $\bar D=D$.
\end{lemma}

\proof See Appendix~\ref{app:lemma-stability}.\hfill\QED
Lemma~\ref{lemma:stability} proves that the simple constraint~\eqref{eq:stability-constraint} does not limit the expressivity of the linear model architecture.
In our experiments, we relax~\eqref{eq:stability-constraint} by including the additional penalty
\begin{equation}
    r_A(z)=\rho_A\max\{\|A\|_2^2-1+\epsilon_A,0\}^2
\label{eq:stability-constraint-penalty}
\end{equation}
in $r(z)$, where $\rho_A\gg 1$ and $0\leq\epsilon_A\ll1$. If the identified matrix $A$ has eigenvalues outside the unit disk, the optimization can be repeated with a larger value of $\rho_A$ and/or $\epsilon_A$. 

\subsection{DC gain}
Assuming a set of steady-state input/output pairs $(y^{\rm ss}_{j},u^{\rm ss}_{j})$ is also available, 
$j=1,\ldots,S$, we can improve the DC gain of the identified model by including the additional regularization
term
\begin{equation}
    \begin{aligned}
    r_s(z)&=\frac{\rho_s}{S}\sum_{j=1}^S\|y^{\rm ss}_j-Cx^{\rm ss}_j-Du^{\rm ss}_j-f_y(x^{\rm ss}_j,u^{\rm ss}_j;\ty)\|_2^2\\
    x^{\rm ss}_j&=Ax^{\rm ss}_j+Bu^{\rm ss}_j+f_x(x^{\rm ss}_j,u^{\rm ss}_j;\tx),\ j=1,\ldots,S.
    \end{aligned}
\label{eq:DCgain}
\end{equation}
The steady-state state $x^{\rm ss}_j$ can be evaluated while optimizing $z$
by an automatically differentiable solver, such as a linear system solver 
in the case of linear models~\eqref{eq:lin-sysid}, or, more generally, a nonlinear
solver, e.g., the limited-memory Broyden solver~\cite{Bro65}. 

\section{Non-smooth nonlinear optimization}
\label{sec:nlp}
When both $f$ and $r$ are smooth functions in~\eqref{eq:training-condensed},
the system identification problem can be solved by any general-purpose 
derivative-based NLP solver. The presence of $\ell_1$- and group-Lasso regularization
requires a little care. Nonlinear non-smooth optimization solvers 
exist that can deal with such penalties, see, e.g., 
the general purpose non-smooth NLP solvers~\cite{CMO17,BCLOS20,STP17}, 
generalized Gauss Newton methods combined with ADMM~\cite{Bem23e},
and methods specifically introduced for $\ell_1$-regularized problems such as the orthant-wise limited-memory quasi-Newton (OWL-QN) method~\cite{AG07}. {\it Stochastic} gradient descent methods based on mini-batches like Adam~\cite{KB14}
would be quite inefficient to solve~\eqref{eq:training}, due to the difficulty in splitting the learning problem in multiple sub-experiments and the very slow convergence of those methods compared to quasi-Newton methods, as we will show in Section~\ref{sec:results} (see also the comparisons reported in~\cite{Bem23e}).

When dealing with large datasets ($Nn_y\gg 1$) and models ($n_z\gg 1$), solution methods based on Gauss-Newton and Levenberg-Marquardt ideas, like the one suggested in~\cite{Bem23e}, can be inefficient, due to the large size $Nn_y\times n_z$ of the required \emph{Jacobian} matrices of the residuals. In such situations, L-BFGS approaches~\cite{LN89} combined with efficient
automatic differentiation methods can be more effective, as processing one epoch of data requires evaluating the \emph{gradient} $\nabla_z f$ of the loss function plus $\nabla_z r$.

We show now how the L-BFGS-B algorithm~\cite{BLNZ95}, an extension of
L-BFGS to handle bound constraints, can be used to also handle $\ell_1$-
and group-Lasso regularization. To this end, we provide two simple technical results in the remainder of this section.

\begin{lemma}
\label{lemma:L1-reg}
Consider the $\ell_1$-regularized nonlinear programming problem
\begin{equation}
    \min_x f(x)+\tau\|x\|_1+r(x)
\label{eq:x-prob}
\end{equation}
where $f:\rr^{n}\to\rr$, $\tau>0$, and function $r:\rr^n\to\rr$ is such that $r(x)=\sum_{i=1}^nr_i(x_i)$.
In addition, let functions $r_i:\rr\to\rr$ be convex and positive semidefinite ($r_i(x_i)\geq 0$, $\forall x_i\geq 0$, $r_i(0)=0$). Let $g:\rr^n\times\rr^n\to\rr$ be defined by
\begin{equation}
    g(y,z) = f(y-z)+\tau[1\ \ldots\ 1]\smallmat{y\\z}+r(y)+r(-z)
\label{eq:g(y,z)}
\end{equation}
and consider the following bound-constrained nonlinear program
\begin{equation}
    \min_{y,z\geq 0} g(y,z).
\label{eq:yz-prob}
\end{equation}
Then any solution $y^*,z^*$ of Problem~\eqref{eq:yz-prob}
satisfies the complementarity condition $y_i^*z_i^*=0$, $\forall i=1,\ldots,n$,
and $x^*\eqdef y^*-z^*$ is a solution of Problem~\eqref{eq:x-prob}.
\end{lemma}

\proof See Appendix~\ref{app:lemma-L1-reg}.\hfill\QED

Note that, if functions $r_i$ are symmetric, i.e., $r_i(x_i)=r_i(-x_i)$, $\forall i=1,\ldots,n$,
then we can replace $r(-z)$ with $r(z)$ in~\eqref{eq:g(y,z)}. Hence, by applying Lemma~\ref{lemma:L1-reg} with $r(x)=\rho\|x\|_2^2=\rho(\sum_{i=1}^nx_i^2)$, we get the following corollary:
\begin{corollary}
\label{cor:elastic-net}
Consider the following nonlinear programming problem with elastic-net regularization
\begin{subequations}
\begin{equation}
    \min_x f(x)+\tau\|x\|_1+\rho\|x\|_2^2
    \label{eq:x-elastic-net}%
\end{equation}
where $\rho>0$ and $\tau>0$, and let
\begin{equation}
    \smallmat{y^*\\z^*}\in\arg\min_{y,z\geq 0} f(y-z)+\tau[1\ \ldots\ 1]\smallmat{y\\z}+\rho\left\|\smallmat{y\\z}\right\|_2^2.
\label{eq:yz-elastic-net}
\end{equation}%
\end{subequations}
Then $x^*\eqdef y^*-z^*$ is a solution of Problem~\eqref{eq:x-elastic-net}.
\end{corollary}
Note that~\eqref{eq:yz-elastic-net} includes the squared Euclidean norm $\|y\|_2^2+\|z\|_2^2$ instead of $\|y-z\|_2^2$, as it would result from a mere substitution
of $x\rightarrow y-z$ in both $f$ and $r$. The regularization in~\eqref{eq:yz-elastic-net} provides a positive-definite, rather than only positive-semidefinite, term on $[y'\ z']'$, which in turn leads to better numerical properties of the problem, without changing its solution.

\begin{lemma}
\label{lemma:r(x+y)}
Consider the $\ell_1$-regularized nonlinear program
\begin{equation}
    \min_x f(x)+r(x)+\epsilon\|x\|_1
\label{eq:x-prob-2}
\end{equation}
where $f:\rr^n\to\rr$, and let $r:\rr^n\to\rr$ be convex and
symmetric with respect to each variable, i.e., satisfying
\begin{subequations}
\begin{equation}
r(x-2x_i e_i)=r(x),\ \forall x\in\rr^n
\label{eq:r(x+y)-a}
\end{equation}
where $e_i$ is the $i$th column of the identity matrix, and increasing
on the positive orthant
\begin{equation}
r(x_0+\gamma e_i)\geq r(x_0),\ \forall x_0\in\rr^n,\ x_0\geq 0,\ \gamma\geq 0
\label{eq:r(x+y)-b}
\end{equation}
\label{eq:r(x+y)-conditions}%
\end{subequations}
Let $\epsilon>0$ be an arbitrary (small) number, let
$g:\rr^n\times\rr^n\to\rr$ defined by
\begin{equation}
    g(y,z) = f(y-z)+r(y+z)+\epsilon[1\ \ldots\ 1]\smallmat{y\\z}
\label{eq:g(y,z)-2}
\end{equation}
and consider the following bound-constrained nonlinear program
\begin{equation}
    \min_{y,z\geq 0} g(y,z).
\label{eq:yz-prob-2}
\end{equation}
Then, any solution $y^*,z^*$ of Problem~\eqref{eq:yz-prob-2}
satisfies the complementarity condition $y_i^*z_i^*=0$, $\forall i=1,\ldots,n$,
and $x^*\eqdef y^*-z^*$ is a solution of Problem~\eqref{eq:x-prob-2}.
\end{lemma}

\proof See Appendix~\ref{app:lemma-r(x+y)}.\hfill\QED

\begin{corollary}
\label{cor:grouplasso}
Consider the following nonlinear programming problem with group-Lasso regularization
\begin{subequations}
\begin{equation}
    \min_x f(x)+\epsilon\|x\|_1+\tau_g\sum_{i=1}^{n_g}\|I_ix\|_2
    \label{eq:x-grouplasso}%
\end{equation}
where $\tau_g>0$ and $\epsilon>0$ is an arbitrary small number. Let
\begin{equation}
    \smallmat{y^*\\z^*}\in\arg\min_{y,z\geq 0} f(y-z)+
    \epsilon[1\ \ldots\ 1]\smallmat{y\\z}+\tau_g\sum_{i=1}^{n_g}\|I_i(y+z)\|_2
\label{eq:yz-grouplasso}
\end{equation}%
\end{subequations}
Then $x^*\eqdef y^*-z^*$ is a solution of Problem~\eqref{eq:x-grouplasso}.
\end{corollary}
\proof See Appendix~\ref{app:cor-r(x+y)}.\hfill\QED

Lemma~\ref{lemma:r(x+y)} and Corollary~\ref{cor:grouplasso} enable us to also handle group-Lasso regularization terms by bound-constrained nonlinear optimization in which the objective function admits partial derivatives on the feasible domain. In particular, consider a basis vector $e_j$ such that $I_ie_j\neq 0$. For $r(y,z)=\|I_i(y+z)\|_2$, $\lim_{\alpha\rightarrow 0^+}\frac{r(\alpha e_j,0)-r(0,0)}{\alpha}=\lim_{\alpha\rightarrow 0^+}\frac{r(0, \alpha e_j)-r(0,0)}{\alpha}=\lim_{\alpha\rightarrow 0^+}\frac{|\alpha|}{\alpha}=1$, where the limit $\alpha\rightarrow 0^+$ is taken since $y,z\geq 0$; on the other hand, for $r(x)=\|I_ix\|_2$, $\lim_{\alpha\rightarrow 0}\frac{r(\alpha e_j)-r(0)}{\alpha}=\lim_{\alpha\rightarrow 0}\frac{|\alpha|}{\alpha}$ is not defined, where the limit $\alpha\rightarrow 0$ is taken
 since there are no sign restrictions on the components of $x$.

Finally, it is easy to prove that Lemma~\ref{lemma:L1-reg} and Lemma~\ref{lemma:r(x+y)}
extend to the case in which only a subvector $x_I$ of $x$ enters
the non-smooth regularization term, $x_I\in\rr^{n_I}$, $n_I<n$. In this case,
we only need to replace $x_I=y_I-z_I$, $y_I,z_I\in\rr^{n_I}$, $y_I,z_I\geq0$.

\subsection{Constraints on model parameters and initial states}
Certain model structures require introducing lower and/or upper bounds
on the parameters defining the model and, possibly, on the initial states. 
For example, as mentioned in Section~\ref{sec:lin-sysid}, positive linear systems require
that the entries of $A,B,C,D$ are nonnegative; similarly, \emph{input-convex neural networks}
require that the weight matrices are nonnegative~\cite{AXK17}. 
General box constraints $x_{\rm min}\leq x\leq x_{\rm max}$, where $x\in\rr^n$ is the optimization vector, can be immediately enforced in L-BFGS-B as bound constraints. When splitting $x=y-z$ to handle $\ell_1$ and group-Lasso regularization, if $x_{\rm max,i}>0$ we can bound the corresponding positive part $0\leq y_i\leq x_{\rm max,i}$, or, if $x_{\rm max,i}<0$, constrain the negative part $z_i\geq -x_{\rm max,i}$
and remove $y_i$ from the optimization vector. Similarly, if $x_{\rm min,i}<0$ one can limit the corresponding negative part by $0\leq z_i\leq -x_{\rm min,i}$,
or, if $x_{\rm min,i}>0$, constrain $y_i\geq -x_{\rm min,i}$ and remove $z_i$.

General constraints $h_I(z)\leq 0$, $h_E(z)=0$, $h:\rr^{n_z}\to\rr^{n_{hI}}$, 
$h:\rr^{n_z}\to\rr^{n_{hE}}$, as shown in~\eqref{eq:stability-constraint-penalty}
for stability constraints, can be addressed in the learning problem~\eqref{eq:cost-reg} via penalty 
functions, such as 
\begin{equation}
    \min_z f(z)+r(z)+\rho_h\left(\sum_{i=1}^{n_{hI}}\max\{h_{Ii}(z),0\}^2+\sum_{j=1}^{n_{hE}}h_{Ej}(z)^2\right)
\end{equation}
where $\rho_h>0$ is a (large) penalty parameter. 

\subsection{Preventing numerical overflows}
Solving~\eqref{eq:training} directly from an arbitrary initial condition $z_0$
can lead to numerical instabilities during the initial optimization steps. 
During training, to prevent such an effect, when evaluating the loss $f(z)$ we saturate the state vector
$x_{k+1}=\textrm{sat}(Ax_k+Bu_k+f_x(x_k,u_k;\tx),x_{\rm sat})$,
where $\textrm{sat}(x,x_{\rm sat})=\min\{\max\{x,-x_{\rm sat}\},x_{\rm sat}\}$,
vector $x_{\rm sat}\in\rr^n$, and the $\min$ and $\max$ functions are applied componentwise. 
As minimizing the open-loop simulation error $f(z)$ in~\eqref{eq:training-cost-condensed} and regularizing $x_0$ and the model coefficients discourage the presence of unstable dynamics, in most cases the saturation term will not be active at the optimal solution if $x_{\rm sat}$ is chosen large enough. After identifying the model, it is easy to test on training data whether the saturation constraint is inactive, so to prove
the redundancy of the saturation function.

At the price of additional numerical burden, to avoid introducing non-smooth
terms in the problem formulation, the soft-saturation function
$\textrm{sat}_\gamma(x,x_{\rm sat})=x_{\rm sat}+\frac{1}{\gamma}\log\frac{1+e^{-\gamma(x+x_{\rm sat})}}{1+e^{-\gamma(x-x_{\rm sat})}}$
could be used as an alternative to $\textrm{sat}(x,x_{\rm sat})$, where the larger $\gamma>0$ the closer the function is to hard saturation.

\subsection{Initial state reconstruction for validation}
\label{sec:initial-state}
To validate a given trained model on new test data
$\{\bar u(0),\bar y(0),\ldots,\bar u(\bar N-1),\bar y(\bar N-1)\}$,
we need a proper initial condition $\bar x_0$ for 
computing open-loop output predictions. Clearly,
if an initial state encoder as described in~\eqref{eq:x0-encoder} 
was also trained, we can simply set $\bar x_0=f_e(\bar v_0;\theta_e)$.
Otherwise, as we suggested in~\cite{Bem23b}, we can solve the nonlinear optimization problem of dimension $n_x$ 
\begin{equation}
    \min_{\bar x_0} r_x(x_0)+\frac{1}{\bar N}\sum_{k=0}^{\bar N-1}\ell(\bar y(k),\hat y(k))
\label{eq:min-loss-state}
\end{equation}
via, e.g., global optimizers such as Particle Swarm Optimization (PSO), 
where $\hat y(k)$ are generated by iterating~\eqref{eq:model}
from $x(0)=\bar x_0$. 

In this paper, in alternative, we propose to 
run an EKF (forward in time) and Rauch-Tung-Striebel (RTS) smoothing~\cite[p.~268]{SS23} (backward in time), based on the learned model, for $N_e$ times. 
The approach is summarized in Appendix~\ref{app:RTS} for completeness.
Clearly, for further refinement, the initial state obtained by the EKF/RTS pass can be used
as the initial guess of a local optimizer (like L-BFGS) solving~\eqref{eq:min-loss-state}.

\section{Numerical results}
\label{sec:results}
We apply the nonlinear programming formulations described in Section~\ref{sec:nlp} to identify linear state-space models and train recurrent neural networks on synthetic and real-world datasets. All experiments are run in Python 3.11 on an Apple M1 Max machine using JAX~\cite{JAX} for automatic differentiation. The L-BFGS-B solver~\cite{LN89} is used via the JAXopt interface~\cite{JAXopt} to solve bound-constrained nonlinear programs. When using group-Lasso penalties as in~\eqref{eq:yz-grouplasso}, 
we constrain $y,z\geq\epsilon$ to avoid possible numerical issues in JAX while
computing partial derivatives when $y=z=0$, with $\epsilon=10^{-16}$. We apply standard scaling $x\leftarrow (x-\bar x)/\sigma_x$ to each input and output signal, where $\bar x$ and $\sigma_x$ are, respectively, the empirical mean and standard deviation computed on training data. 
When learning linear models, initial states are reconstructed using the EKF+RTS method described in Appendix~\ref{app:RTS}, running a single forward EKF and backward RTS pass ($N_e=1$). Unless stated differently, matrix $A$ is initialized as $0.5I$, the remaining coefficients are initialized by drawing random numbers from the normal distribution with standard deviation 0.1.
When running Adam as a gradient-descent method, due to the absence of line search,
we store the model corresponding to the best loss found during the iterations.

For single-output systems, the quality of fit is measured by the classical $R^2$-score,
$R^2 = 100\left(1-\frac{\sum_{k=1}^N(y_{k}-\hat y_{k})^2}
        {\sum_{k=1}^N(y_{k}-\frac{1}{N}\sum_{i=1}^N y_{k})^2}\right)$,
where $\hat y_{k}$ is the output simulated in open-loop from $x_0$. 
In the case of multiple outputs, we consider the average $R^2$-score
$\overline{R^2}\triangleq \frac{1}{n_y}\sum_{i=1}^{n_y}R^2_i$, 
where $R^2_i$ is the $R^2$-score measuring the quality of fit of the $i$th component of the output. 

\subsection{Identification of linear models}

\subsubsection{Cascaded-Tanks benchmark}
We consider the single-input/single-output Cascaded-Tanks benchmark problem described in~\cite{SMWN16}. 
The dataset consists of 1024 training and 1024 test standard-scaled samples. L-BFGS-B 
is run for a maximum of 1000 function evaluations, after being initialized by 1000 Adam iterations. Due to the lack of line search, which may reduce the step size considerably and trap the algorithm into a local minimum, we found that Adam is a good way to initially explore the optimization-vector space and provide a good initial guess to L-BFGS for refining the solution, without suffering from the slow convergence of gradient descent methods.
We select the model with the best $R^2$ on training data obtained out of 5 runs
from different initial conditions. We also run the N4SID method implemented in
Sippy~\cite{AVDBP18} (\texttt{sippy}), and the \texttt{n4sid} method (with both
focus on prediction and simulation) and \texttt{ssest} 
(prediction error method, with focus on simulation) of the System Identification Toolbox for MATLAB (\texttt{M}), all with stability enforcement enabled. 
Table~\ref{tab:cascaded-tanks} summarizes the obtained results. The average CPU time spent per run is about 2.4~s (\texttt{lbfgs}), 30~ms (\texttt{sippy}), 50~ms (\texttt{n4sid} with focus on prediction), 0.3~s (\texttt{n4sid} with focus on simulation), 0.5~s (\texttt{ssest}).

It is apparent from the table that, while our approach always provides reliable results,
the other methods fail in some cases. We observed that the main reason for failure
is due to numerical instabilities of the N4SID method (used by \texttt{n4sid}
and, for initialization, by \texttt{ssest}). 

\begin{table}
\begin{center}
{\footnotesize
\setlength{\tabcolsep}{6pt}
\begin{tabular}{c|ccc|cccl}
 & \multicolumn{3}{|c|}{$R^2$ (training)} & \multicolumn{3}{|c}{$R^2$ (test)} \\
$n_x$\hspace*{-.5em} & \texttt{lbfgs} & \texttt{sippy}& \texttt{M} & \texttt{lbfgs} & \texttt{sippy} & \texttt{M} &\\\hline
1\hspace*{-.5em} &  87.43 &  56.24 &  87.06 & 83.22 &  52.38 &  83.18 &\hspace*{-1em}(\texttt{ssest})\hspace*{-.5em}\\
2\hspace*{-.5em} &  94.07 &  28.97 &  93.81 & 92.16 &  23.70 &  92.17 &\hspace*{-1em}(\texttt{ssest})\hspace*{-.5em}\\
3\hspace*{-.5em} &  94.07 &  74.09 &  93.63 & 92.16 &  68.74 &  91.56 &\hspace*{-1em}(\texttt{ssest})\hspace*{-.5em}\\
4\hspace*{-.5em} &  94.07 &  48.34 &  92.34 & 92.16 &  45.50 &  90.33 &\hspace*{-1em}(\texttt{ssest})\hspace*{-.5em}\\
5\hspace*{-.5em} &  94.07 &  90.70 &  93.40 & 92.16 &  89.51 &  80.22 &\hspace*{-1em}(\texttt{ssest})\hspace*{-.5em}\\
6\hspace*{-.5em} &  94.07 &  94.00 &  93.99 & 92.17 &  92.32 &  88.49 &\hspace*{-1em}(\texttt{n4sid})\hspace*{-.5em}\\
7\hspace*{-.5em} &  94.07 &  92.47 &  93.82 & 92.17 &  90.81 & < 0 &\hspace*{-1em}(\texttt{ssest})\hspace*{-.5em}\\
8\hspace*{-.5em} &  94.49 & < 0 &  94.00 & 89.49 & < 0 & < 0 &\hspace*{-1em}(\texttt{n4sid})\hspace*{-.5em}\\
9\hspace*{-.5em} &  94.07 & < 0 & < 0 & 92.17 & < 0 & < 0 &\hspace*{-1em}(\texttt{ssest})\hspace*{-.5em}\\
10\hspace*{-.5em} &  94.08 &  93.39 & < 0 & 92.17 &  92.35 & < 0 &\hspace*{-1em}(\texttt{ssest})\hspace*{-.5em}\\
\hline
\end{tabular}}
\end{center}
\caption{Cascaded-tanks benchmark: $R^2$-scores on training and test data obtained
by running Adam followed by L-BFGS from 5 different initial conditions
and selecting the model with best $R^2$-score on training data (\texttt{lbfgs}), by Sippy~\cite{AVDBP18} (\texttt{sippy}),
and by the System Identification Toolbox for MATLAB, selecting
the best model obtained by running both \texttt{n4sid} and \texttt{ssest}.}
\label{tab:cascaded-tanks}
\end{table}

\subsubsection{Group-Lasso regularization for model-order reduction}
To illustrate the effectiveness of group-Lasso regularization in reducing the model order, we generated
2000 training data from the following linear system
\beqarno
   x_{k+1}\hspace*{-1.1em}&=&\hspace*{-1.1em}\smallmat{
0.96 & 0.26 & 0.04 & 0 & 0 & 0 \\
-0.26 & 0.70 & 0.26 & 0 & 0 & 0 \\
0 & 0 & 0.93 & 0.32 & 0.07 & 0 \\
0 & 0 & -0.32 & 0.61 & 0.32 & 0 \\
0 & 0 & 0 & 0 & 0.90 & 0.38 \\
0 & 0 & 0 & 0 & -0.38 & 0.52}\hspace*{-0.2em}x_k+\hspace*{-0.2em}
\smallmat{
0 & 0 \\
0 & 0 \\
0.07 & 0 \\
0.32 & 0 \\
0 & 0.10 \\
0 & 0.38}\hspace*{-0.2em}u_k\hspace*{-0.2em}+\hspace*{-0.2em}\xi_k\\
y_k\hspace*{-1.1em}&=&\hspace*{-1.1em}\smallmat{x_1\\x_3}\hspace*{-0.2em}+\hspace*{-0.2em}\eta_k
\eeqarno
where $\xi_k$ and $\eta_k$ are vectors whose entries are zero-mean Gaussian independent noise signals with standard deviation 0.01.

\begin{figure}[t]
\begin{center}
\includegraphics[width=\hsize]{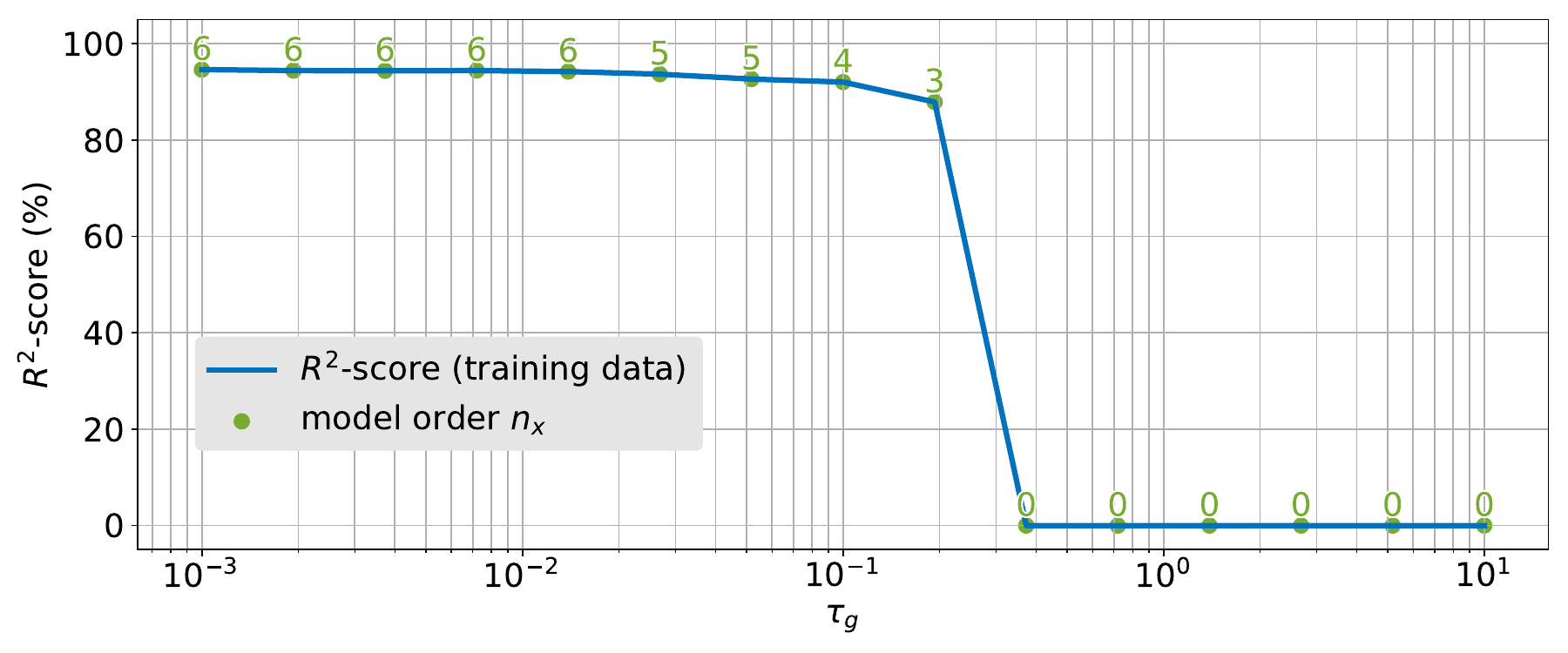}
\end{center}
\caption{$R^2$-score on training data and resulting model order obtained with the group-Lasso penalty~\eqref{eq:group-Lasso-x} for different values of $\tau_g$.}
\label{fig:tau-x}
\end{figure}

Figure~\ref{fig:tau-x} shows the $R^2$-score on training data and the corresponding model order obtained by running 1000 Adam iterations followed by at most 1000 L-BFGS-B function evaluations to handle the regularization term~\eqref{eq:group-Lasso-x} for different values of the group-Lasso penalty $\tau_g$. The remaining penalties are
$\rho_\theta=\rho_x=10^{-3}$, and $\tau=\epsilon=10^{-16}$. For each value of $\tau_g$,
the best model in terms of $R^2$-score on training data is selected out of 10 runs from different initial conditions. The figure clearly shows the effect of $\tau_g$ in trading off the quality of fit and the model order. The CPU time is about 3.85~s per run.

\subsubsection{Group-Lasso regularization for input selection}
To illustrate the effectiveness of group-Lasso regularization in reducing the number of inputs 
of the system, we generate 10000 training data from a randomly generated stable linear system with 3 states, one output, 10 inputs, and zero-mean Gaussian process and measurement noise with standard deviation 0.01. The last 5 columns of the $B$ matrix are divided by 1000 to make the last 5 inputs almost redundant. 

\begin{figure}[t]
\begin{center}
\includegraphics[width=\hsize]{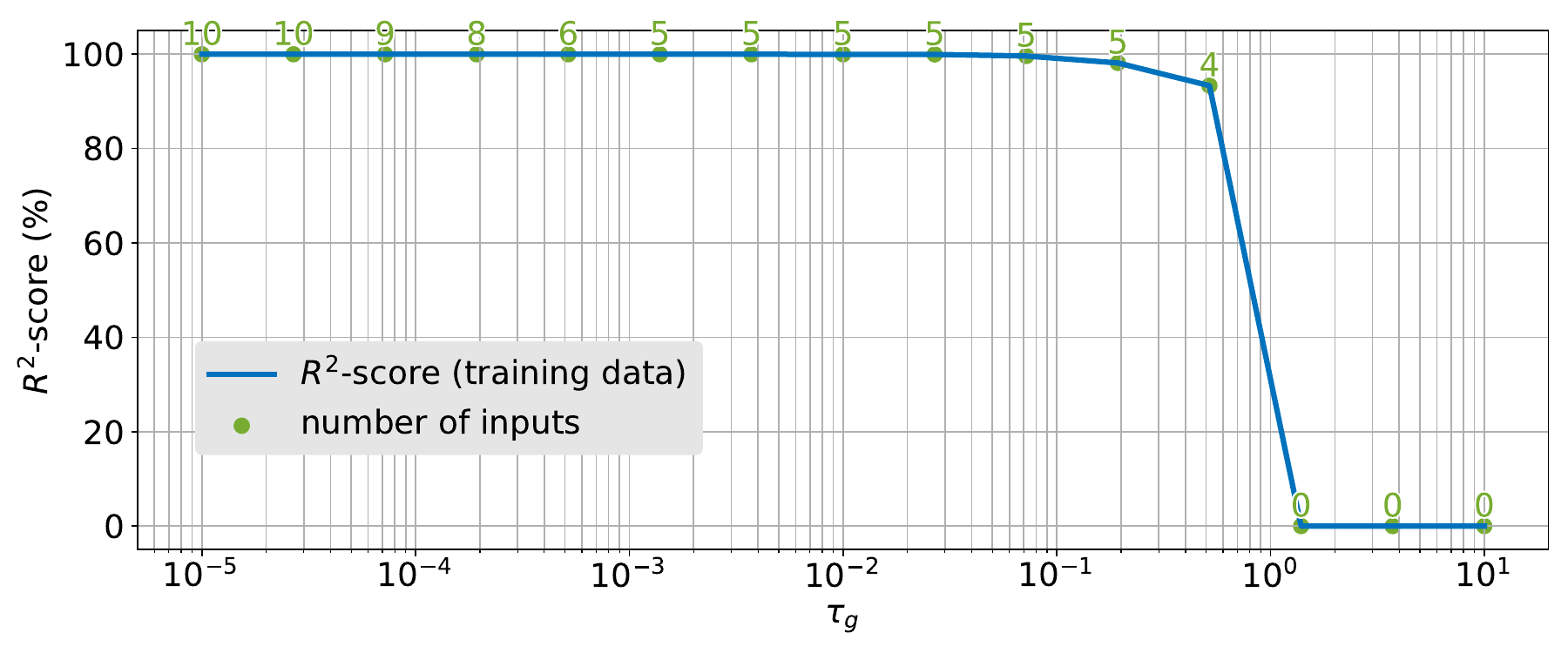}
\end{center}
\caption{$R^2$-score on training data and resulting number of model inputs under the group-Lasso penalty~\eqref{eq:group-Lasso-u} for different values of $\tau_g$.}
\label{fig:tau-u}
\end{figure}

Figure~\ref{fig:tau-u} shows the $R^2$-score on training data and the corresponding number of
nonzero columns in the $B$ matrix obtained by running 1000 Adam iterations followed by a maximum of 1000 L-BFGS-B function evaluations to handle the term~\eqref{eq:group-Lasso-u} with $\rho_\theta=\rho_x=10^{-8}$, $\tau=\epsilon=10^{-16}$, and different values of the group-Lasso penalty $\tau_g$. For each value of $\tau_g$,
the best model in terms of $R^2$-score on training data is selected out of 10 runs from different initial conditions. The figure clearly shows the effect of $\tau_g$ in trading off the quality of fit and the number of inputs kept in the model. The CPU time is about 3.71~s per run.

\subsection{Stability constraint}
\label{sec:stability-example}
To test the effectiveness of the approach to enforce stability via~\eqref{eq:stability-constraint}, we generated 1000 training and test data from the following sightly unstable linear system
\[
    \ba{rcl}
    x_{k+1}&=&\smallmat{1.0001& 0.5& 0.5\\0& 0.9& -.2\\0&0&0.7}x_k+\smallmat{-0.4168\\-0.0563\\-2.1362}u_k+\xi_k\\
    y_k&=&\smallmat{1.6403 & -1.7934& -0.8417}x_k+\eta_k
    \ea
\]
where $\xi_k\sim\NN(0,0.01^2I)$, $\eta\sim\NN(0,0.05^2)$. The training problem is solved in 6.4~s under the additional penalty~\eqref{eq:stability-constraint-penalty} with $\rho_A=10^3$ and $\epsilon_A=10^{-3}$. The obtained $R^2$-score is 92.27 on training data and 91.41 on test data, the eigenvalues of the identified $A$ matrix are $0.57871687$, $0.99996862$, and $0.92904764$.

\subsection{Industrial robot benchmark}
\label{sec:robot}
\subsubsection{Nonlinear system identification problem}
The dataset described in~\cite{WGUR22} contains input and output samples collected from KUKA KR300 R2500 ultra SE industrial robot movements, resampled at 10 Hz, where the input $u\in\rr^6$ collects motor torques (Nm) and the output $y\in\rr^6$ joint angles (deg). The dataset contains two experimental
traces: a training dataset of $N=39988$ samples and a test dataset of $N_t=3636$ samples. 
Our aim is to obtain a discrete-time recurrent neural network (RNN) model in residual form~\eqref{eq:model},
with sample time $T_s=100$~ms, by minimizing the mean squared error (MSE) between the measured output $y_k$ and the open-loop prediction $\hat y_{k}$ obtained by simulating~\eqref{eq:model}. Standard scaling is applied on both input and output signals.

The industrial robot identification benchmark introduces multiple challenges, as the system generating the data is highly nonlinear, multi-input multi-output, data are slightly over-sampled (i.e., $\|y_k-y_{k-1}\|$ is often very small), and the training dataset contains a large number of samples, that complicates solving the training problem. As a result, minimizing the open-loop simulation error on training data is rather challenging from a computational viewpoint.
We select the 
model order $n_x=12$ and
shallow neural networks $f_x$, $f_y$ with, respectively, 
36 and 24 neurons, and swish activation function $\frac{x}{1+e^{-x}}$.

We first identify matrices $A$, $B$, $C$ and a corresponding
initial state $x_0$ on training data by solving problem~\eqref{eq:lin-sysid} with 
$\rho_\theta=\rho_x=0.001$, $\tau=0$ in~\eqref{eq:l2-l1-regularization}, which takes $9.12~s$ by running 1000 L-BFGS-B function evaluations, starting from the
initial guess $A=0.99I$ and with all the entries of $B$ and $C$ 
normally distributed with zero mean and standard deviation equal to $0.1$. The largest absolute value of the eigenvalues of matrix $A$ is approximately $0.9776$.

The average $R^2$-score on all outputs is $\overline{R^2}$ = 48.2789 on training data and $\overline{R^2}$ = 43.8573
on test data. For training data, the initial state $x_0$ is taken from the optimal 
solution of the NLP problem, while for test data we reconstruct
$x_0$ by running EKF+RTS based on the obtained
model for $N_e=10$ epochs.

For comparison, we also trained the same model structure using the 
N4SID method~\cite{VD94} implemented in the System Identification Toolbox for MATLAB R2023b \cite{Lju01} 
with focus on open-loop simulation. This took $36.21$~s on the same machine
and provided the lower-quality result $\overline{R^2}$ = 39.2822 on training data and $\overline{R^2}$ = 32.0410 on test data. The N4SID algorithm in \texttt{sippy} with default options 
did not succeed in providing meaningful results on the training dataset.

After fixing $A$, $B$, $C$, we train $\tx,\ty,x_0$ by minimizing the MSE open-loop prediction loss under the elastic net regularization~\eqref{eq:l2-l1-regularization} to limit overfitting and possibly reduce the number of nonzero entries in $\tx,\ty$, with regularization coefficients $\rho_\theta=0.01$ and $\rho_x=0.001$ in~\eqref{eq:l2-l1-regularization}, and different values of
$\tau$. As a result, the total number of variables to optimize is 1650, i.e., $\dim(\tx)+\dim(\ty)=1590$ 
plus 12 components of the initial state $x_0$.
This amounts to $2\cdot 1590+12=3192$ optimization variables when applying the method of 
Lemma~\ref{lemma:L1-reg}. We achieved the best results by 
randomly sampling 100 
initial conditions for the model parameters and picking
up the best one in terms of loss $f(z)$ as defined in~\eqref{eq:training-cost-condensed}
before running Adam for 2000 iterations; then, we 
refined the solution by running a maximum of 2000 L-BFGS-B function evaluations. For a given value of $\tau$, solving the training problem took an average of 92~ms per iteration on a single core of the CPU. Solving directly the same $\ell_1$-regularized nonlinear programming problem as in~\eqref{eq:x-prob}
via Adam~\cite{KB14} (1590 variables) with constant learning rate $\eta=0.01$
took about 66~ms per iteration, where $\eta$ is chosen to have a good tradeoff 
between the expected decrease and the variance of the function values generated by the optimizer.

\begin{figure}[t]
\begin{center}
\includegraphics[width=\hsize]{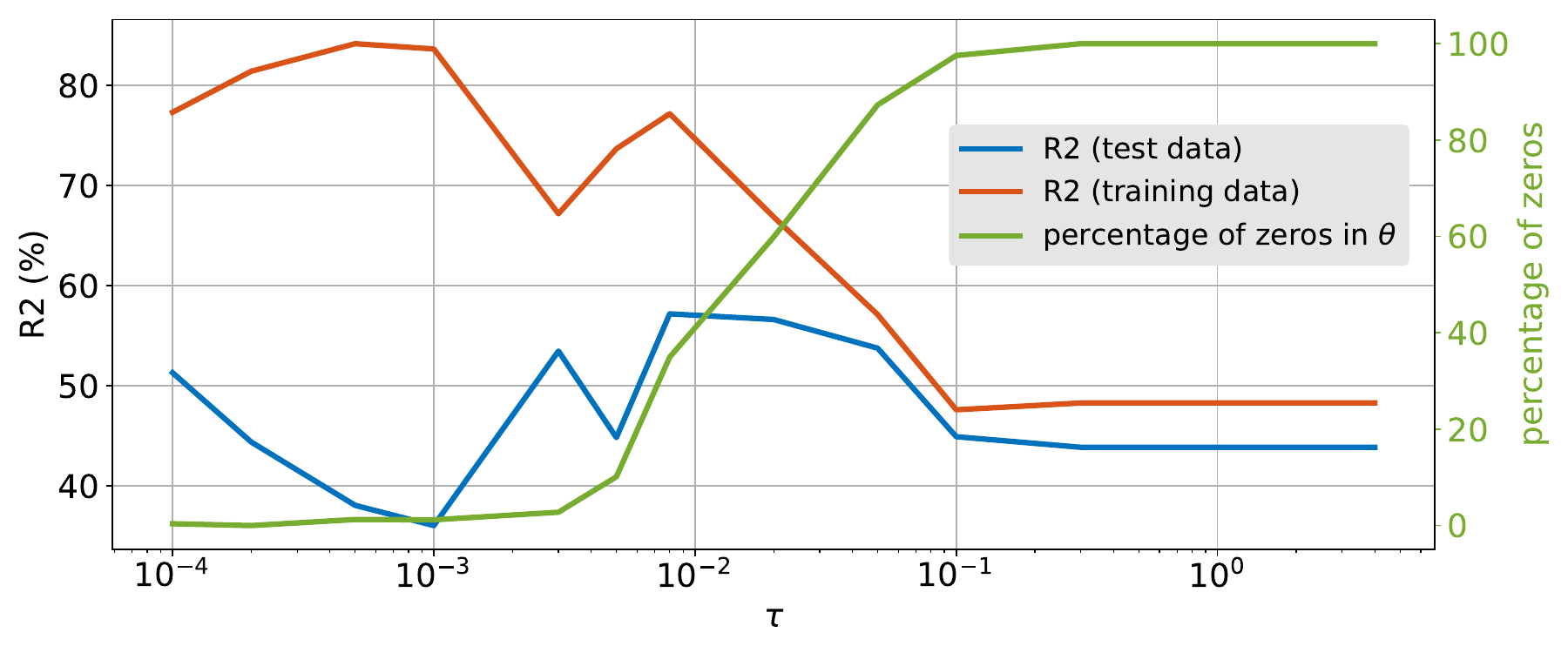}
\end{center}
\caption{Average $R^2$-score of open-loop predictions on training/test data and model sparsity, as a function of $\tau$. For each value of $\tau$, we show the best $\overline{R^2}$-score obtained on test data out of 30 runs and the corresponding $\overline{R^2}$-score on training data and model sparsity.}
\label{fig:R2-tau}
\end{figure}

The average $R^2$-score ($\overline{R^2}$) as a function of the $\ell_1$-regularization parameter $\tau$ is shown in Figure~\ref{fig:R2-tau}.
For each value of $\tau$, we identified a model and its initial state starting from 30 different initial best-cost values of the model parameters and $x_0=0$, and the model with best $\overline{R^2}$ on test data is plotted for each $\tau$.
It is apparent from the figure that the better the corresponding $\overline{R^2}$ is on training data, the worse it is on test data; this denotes that the training dataset is not informative enough, as overfitting occurs.

The model obtained that leads to the best $\overline{R^2}$ on test data
corresponds to setting $\tau$=0.008. Table~\ref{tab:R2} shows the 
resulting $R^2$-scores obtained by running such a model in open-loop simulation for each output, along with the scores obtained by running the linear model $(A,B,C)$. The
latter coincide with the $R^2$-scores shown in Figure~\ref{fig:R2-tau}
for large values of $\tau$, which lead to $\tx=0$, $\ty=0$.

\begin{table}[h!]{
\begin{center}\footnotesize
\setlength{\tabcolsep}{4pt}
\begin{tabular}{c|cc|cc}
 & $R^2$ (training) & $R^2$ (test) & $R^2$ (training) & $R^2$ (test) \\
 & RNN model & RNN model & linear model & linear model \\\hline
$y_1$ & 86.0482 & 68.6886 & 63.0454& 63.9364\\ 
$y_2$ & 77.4169 & 70.5481 & 53.1470& 35.2374\\
$y_3$ & 72.3625 & 64.9590 & 63.7287& 55.7936\\
$y_4$ & 75.7727 & 36.4175 & 29.9444& 27.2043\\
$y_5$ & 65.1283 & 32.0540 & 35.4554& 44.2490\\
$y_6$ & 86.1674 & 70.4031 & 44.3523& 36.7233\\\hline
average 
& 77.1493 &  57.1784 &48.2789& 43.8573\\\hline
\end{tabular}
\end{center}}
\caption{Open-loop simulation: $R^2$-scores ($\tau$=0.008).}
\label{tab:R2}
\end{table}

To assess the benefits introduced by running L-BFGS-B, Table~\ref{tab:adam_vs_lbgfsb}
compares the results obtained by running: ($i$) 2000 Adam iterations followed 
by 2000 L-BFGS-B function evaluations, ($ii$) 2000 OWL-QN function evaluations, ($iii$) 
only Adam for 6000 iterations, which is long enough to achieve a similar quality of fit, and
($iv$) 4000 L-BFGS-B iterations without first warm-starting with Adam (``No-Adam'').
The table shows the best result obtained out of 10 runs
in terms of the largest 
$\bar R^2$ on test data. Model coefficients
are considered zero if the absolute value is less than $10^{-6}$.
It is apparent that L-BFGS-B or OWL-QN iterations, compared to pure Adam, 
better minimize the training loss and lead to much sparser models,
providing similar results in terms of both model 
sparsity and training loss, with slightly different tradeoffs.
We remark that, however, OWL-QN would not be capable of handling group-Lasso penalties~\eqref{eq:group-Lasso}. It also evident the benefit of warm-starting L-BFGS-B with Adam.

\begin{figure}[t]
\begin{center}
\includegraphics[width=\hsize]{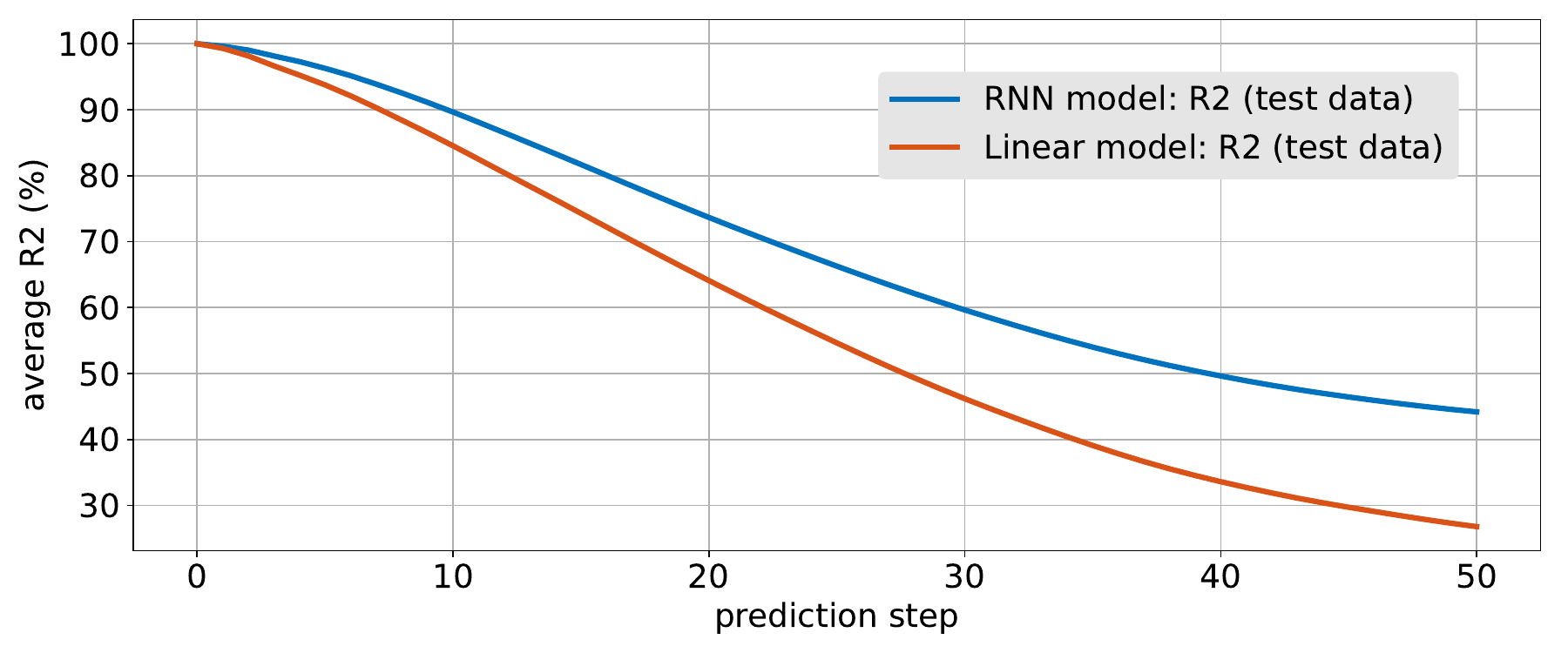}
\end{center}
\caption{Average R$^2$-score of open-loop predictions on test data
from state estimates $\hat x_{k|k}$ obtained by EKF ($\tau$=0.008).}
\label{fig:R2-EKF}
\end{figure}

Finally, we run an extended Kalman filter based on model~\eqref{eq:model}
(with $\tx,\ty$ obtained with $\tau$=0.008) 
and, for comparison, on the linear model $(A,B,C)$ to estimate the hidden states $\hat x_{k|k}$. In both cases, 
as routinely done in MPC practice~\cite{PGA15}, we computed EKF estimates on the model
augmented by the output disturbance model $q_{k+1}=q_k$ plus white noise, with $q_k\in\rr^6$.
The average $R^2$-score of the $p$-step ahead predictions $\hat y_{k+p|k}$
are shown in Figure~\ref{fig:R2-EKF}. This is a more relevant indicator of model quality for MPC purposes than the open-loop simulation error $\hat y_{k|0}-y_k$ considered in Figure~\ref{fig:R2-tau}. 

\begin{table}[h!]{\footnotesize
\centerline{
\setlength{\tabcolsep}{4pt}
\begin{tabular}{l|r|r|r|r|r|r}
         &  adam   &  fcn& $\overline{R^2}$ & $\overline{R^2}$ &\# zeros & CPU\\
solver   &  iters  &  evals & training         & test             & $(\tx,\ty)$ & time (s)\\\hline
L-BFGS-B & 2000  & 2000     & 77.1493& 57.1784& 556/1590 & 309.87\\
OWL-QN   & 2000  & 2000     & 74.7816& 54.0531 & 736/1590 & 449.17\\
Adam     & 6000  & 0        & 71.0687& 54.3636 &   1/1590 & 389.39\\
No-Adam  & 0     & 4000     & 66.8804& 55.6671 & 302/1590 & 1361.08\\\hline
\end{tabular}}}
\caption{Training results obtained by running L-BFGS-B or OWL-QN iterations after Adam iterations,
just Adam, and without running Adam first ($\tau$=0.008). The table shows the result out of 10 runs 
corresponding to the best $R_2$-score achieved on test data.
}
\label{tab:adam_vs_lbgfsb}
\end{table}

\section{Conclusions}
The proposed approach based on bound-constrained nonlinear programming can solve a very broad class of nonlinear system identification problems, possibly under $\ell_1$- and group-Lasso regularization. It
is very general and can handle a wide range of smooth loss and regularization terms, bounds on model coefficients, and model structures. As shown in numerical experiments, the approach is also valid for the identification of linear models, as it seems more stable from a numerical point of view and often provides better results than classical linear subspace methods, possibly at the price of higher computations and the need to rerun the problem from different initial guesses due to local minima.

\newcommand{\noopsort}[1]{}

\appendix
\section{Appendix}

\subsection{Proof of Lemma~\ref{lemma:stability}}
\label{app:lemma-stability}
Since $\Sigma$ is asymptotically stable, there exist symmetric and positive definite matrices $P,Q\in\rr^{n_x\times n_x}$ such that the Lyapunov equation $P-A^TPA=Q$ is satisfied.
Let $T'T=P$ be a Cholesky decomposition of $P$. Then $T^{-T}(T'T-A'T'TA)T^{-1}=\bar Q$,
where $\bar Q\eqdef T^{-T}QT^{-1}\succ 0$. Hence, $I-\bar A'\bar A=\bar Q$, and therefore $x'x-x'\bar A'\bar A x=x'\bar Qx\geq \alpha\|x\|_2^2$, $\forall x\neq 0$, where $\alpha$ is the smallest eigenvalue of $\bar Q$, $\alpha>0$, or equivalently $1-\|\bar Ax\|_2^2/\|x\|_2^2\leq \alpha$
and hence $\frac{\|\bar Ax\|_2}{\|x\|_2}\leq \sqrt{1-\alpha}$. Therefore, $\|\bar A\|_2=\sup_{\|x\|_2\neq 0}\frac{\|\bar A x\|_2}{\|x\|_2}\leq \sqrt{1-\alpha}<1$.
\hfill\QED

\subsection{Proof of Lemma~\ref{lemma:L1-reg}}
\label{app:lemma-L1-reg}
By contradiction, assume that $y_i^*z_i^*\neq 0$ for some index $i$, $1\leq i\leq n$.
Since $y_i^*,z_i^*\geq 0$, this implies that $y_i^*z_i^*>0$. Let $\alpha\eqdef\min
\{y_i^*,z_i^*\}$ and set $\bar y\eqdef y^*-\alpha e_i$, $\bar z\eqdef z^*-\alpha e_i$,
where $e_i$ is the $i$th column of the identity matrix and clearly $\alpha>0$.
By setting $\beta_y\eqdef\frac{\alpha}{y_i^*}$, $\beta_z\eqdef\frac{\alpha}{z_i^*}$,
we can rewrite $\bar y=(1-\beta_y)y^*+\beta_y(y^*-y_i^*e_i)$, $\bar z=(1-\beta_z)z^*+\beta_z(z^*-z_i^*e_i)$,
where clearly $\beta_y,\beta_z\in(0,1]$.
Since $r_i$ are convex functions, $r$ is also convex and satisfies Jensen's inequality:
$r(\bar y) =r((1-\beta_y)y^*+\beta_y(y^*-y_i^*e_i))\leq (1-\beta_y)r(y^*)+\beta_yr(y^*-y_i^*e_i)$. 
Since $r$ is separable, $r_i(0)=0$, and $r_i(y_i^*)\geq 0$, we also get
$r(\bar y)\leq r(y^*)-\beta_y(r(y^*)-r(y^*-y_i^*e_i))=r(y^*)-\beta_y(r_i(y^*_i)-r_i(0))\leq r(y^*)$. 
Similarly, $r(-\bar z)$ = $r((1-\beta_z)(-z^*)$+$\beta_z(-z^*+z_i^*e_i))$
$\leq$ $(1-\beta_z)r(-z^*)$+$\beta_zr(-z^*+z_i^*e_i)$ =
$r(-z^*)$-$\beta_z (r_i(-z_i^*)-r_i(0))$$\leq$$r(-z^*)$. 
Then, since $\alpha>0$ and $\tau>0$, we obtain
$g(\bar y,\bar z)=f(\bar y^*-\bar z^*)+\tau[1\ \ldots\ 1]\smallmat{y^*\\z^*}
-2n\tau\alpha+r(\bar y)+r(-\bar z)< f(\bar y^*-\bar z^*)+\tau[1\ \ldots\ 1]\smallmat{y^*\\z^*}+r(y^*)+r(-z^*)=g(y^*,z^*)$.
This contradicts $(y^*,z^*)\in\arg\min$~\eqref{eq:yz-prob} and therefore
the initial assumption $y_i^*z_i^*\neq 0$. 

Now let $x^*\eqdef y^*-z^*$ and assume by contradiction that $x^*\not\in\arg\min$~\eqref{eq:x-prob}. Let $\bar x\in\arg\min$~\eqref{eq:x-prob}
and set $\bar y\eqdef\max\{\bar x,0\}$, $\bar z\eqdef\max\{-\bar x,0\}$. Then,
since $\bar y_i\bar z_i=0$ and $r_i(0)=0$, $\forall i=1,\ldots,n$, we get
$g(\bar y,\bar z)=f(\bar x)+\tau\|\bar x\|_1+\sum_{i=1}^nr_i(\bar y_i)+r_i(-\bar z_i)
=f(\bar x)+\tau\|\bar x\|_1+\sum_{i=1}^nr_i(\bar x_i)
<f(x^*)+\tau\|x^*\|_1+\sum_{i=1}^nr_i(x_i^*)=g(y^*,z^*)$.
This contradicts $(y^*,z^*)\in\arg\min$~\eqref{eq:yz-prob} and hence
the assumption $x^*\not\in\arg\min$~\eqref{eq:x-prob}.
\hfill\QED

\subsection{Proof of Lemma~\ref{lemma:r(x+y)}}
\label{app:lemma-r(x+y)}
Similarly to the proof of Lemma~\ref{lemma:L1-reg}, assume by contradiction that $y_i^*z_i^*\neq 0$ for some index $i$, $1\leq i\leq n$.
Since $y_i^*,z_i^*\geq 0$, this implies $y_i^*z_i^*>0$. Let $\alpha\eqdef\min
\{y_i^*,z_i^*\}$ and set $\bar y\eqdef y^*-\alpha e_i$, $\bar z\eqdef z^*-\alpha e_i$,
where clearly $\alpha>0$.
By setting $\beta\eqdef\min\left\{\frac{\alpha}{y_i^*},\frac{\alpha}{z_i^*}\right\}$,
we can rewrite $\bar y=(1-\beta)y^*+\beta(y^*-y_i^*e_i)$, $\bar z=(1-\beta)z^*+\beta(z^*-z_i^*e_i)$, 
where clearly $\beta\in(0,1]$. Then,
$r(\bar y+\bar z) =r((1-\beta)(y^*+z^*)+\beta(y^*+z^*-(y_i^*+z_i^*)e_i))
\leq (1-\beta_y)r(y^*+z^*)+\beta r(y^*+z^*-(y_i^*+z_i^*)e_i)\leq r(y^*+z^*)$,
where the first inequality follows from Jensen's inequality due to the convexity of $r$
and the second inequality from~\eqref{eq:r(x+y)-b}.
Then, since $\alpha>0$, we obtain
$g(\bar y,\bar z)=f(\bar y^*-\bar z^*)+r(\bar y+\bar z)+\epsilon[1\ \ldots\ 1]\smallmat{y^*\\z^*}-2n\epsilon\alpha< f(\bar y^*-\bar z^*)+ r(y^*+z^*)+\epsilon[1\ \ldots\ 1]\smallmat{y^*\\z^*}=g(y^*,z^*)$.
This contradicts $(y^*,z^*)\in\arg\min$~\eqref{eq:yz-prob-2} and therefore
the initial assumption $y_i^*z_i^*\neq 0$. 

Now let $x^*\eqdef y^*-z^*$ and assume by contradiction that $x^*\not\in\arg\min$~\eqref{eq:x-prob}. Let $\bar x\in\arg\min$~\eqref{eq:x-prob}
and set $\bar y\eqdef\max\{\bar x,0\}$, $\bar z\eqdef\max\{-\bar x,0\}$. Then,
since $\bar y_i\bar z_i=0$, $\forall i=1,\ldots,n$, we get
$g(\bar y,\bar z)=f(\bar y-\bar z)+r(\bar y+\bar z)+\epsilon[1\ \ldots\ 1]\smallmat{\bar y\\\bar z}=f(\bar x)+r(\bar x)+\epsilon\|\bar x\|_1<f(x^*)+r(x^*)+\epsilon\|x^*\|_1=f(y^*-z^*)+r(y^*-z^*)+\epsilon[1\ \ldots\ 1]\smallmat{y^*\\\bar z^*}=f(y^*-z^*)+r(y^*+z^*)+\epsilon[1\ \ldots\ 1]\smallmat{y^*\\\bar z^*}$, 
where the equivalences $r(\bar y+\bar z)=r(\bar y-\bar z)=r(\bar x)$ and
$r(y^*+z^*)=r(y^*-z^*)=r(x^*)$ follow by~\eqref{eq:r(x+y)-a} 
and the complimentarity conditions $\bar y_i\bar z_i=0$, $y^*_iz^*_i=0$, respectively.
This contradicts $(y^*,z^*)\in\arg\min$~\eqref{eq:yz-prob} and hence
the assumption $x^*\not\in\arg\min$~\eqref{eq:x-prob}.
\hfill\QED

\subsection{Proof of Corollary~\ref{cor:grouplasso}}
\label{app:cor-r(x+y)}
By the triangle inequality, we have that
$\|I_i(\alpha x_1+(1-\alpha)x_2)\|_2\leq \|\alpha I_ix_1\|_2 + \|(1-\alpha)I_ix_2\|_2=\alpha \|I_ix_1\|_2 + (1-\alpha)\|I_ix_2\|_2$, for all $0\leq \alpha\leq 1$ and $i=1,\ldots,n_g$. Hence,
$g(x)=\sum_{i=1}^{n_x}\|I_ix\|_2$ satisfies Jensen's inequality
and is therefore convex. Moreover, since $\|I_ix\|_2^2=\sum_{j\in J_i}x_j^2$,
where $J_i$ is the set of indices of the components of $x$ corresponding to the $i$th group, 
$g(x)$ is clearly symmetric with respect to each variable $x_j$. Finally, for all vectors 
$x_0\geq 0$ and $\gamma\geq 0$, we clearly have that $x_{0j}+\gamma\geq x_{0j}$,
and therefore $\|I_i(x_{0}+\gamma e_j)\|_2\geq \|I_ix_{0}\|_2$,
proving that $g$ is also increasing on the positive orthant. 
Hence, the corollary follows by applying Lemma~\ref{lemma:r(x+y)}.\hfill\QED

\subsection{EKF and RTS smoother for initial state reconstruction}
\label{app:RTS} 
Algorithm~\ref{algo:RTS} reports the procedure used to estimate the initial state $x_0$
of a generic nonlinear parametric model
\begin{equation}
    \begin{split}
        x_{k+1}&=f(x_k,u_k;\theta)\\
        \hat y_k&=g(x_k,u_k;\theta)
    \end{split}
    \label{eq:nlmodel}
\end{equation}
for a given input/output dataset $(u_0,y_0)$, $\ldots$, $(u_{N-1},y_{N-1})$ based on multiple
runs of EKF with Joseph stabilized covariance updates~\cite[Eq.~(4.26)]{BJ68}
and RTS smoothing, where $\theta$ is the learned vector of parameters. 

The procedure requires storing $\{x_{k|k},x_{k+1|k},P_{k|k},P_{k+1|k}\}_{k=0}^{N-1}$ and, in the case
of nonlinear models, also the Jacobian matrices $A_k\eqdef\frac{\partial f(x_{k|k},u_k;\theta)}{\partial x}$
to avoid recomputing them twice. 
In the reported examples
we used $x_0^s=0$, $P_0^s=\frac{1}{\rho_x N}I$, which is equivalent to
the regularization term  $\frac{\rho_x}{2}\|x_0\|_2^2$ (cf. Eq.~(12) in~\cite{Bem23b}), 
and we set $Q=10^{-8}I$, $R=I$.

\begin{algorithm}[h!]
    \caption{Initial state reconstruction by EKF and RTS smoothing}
    \label{algo:RTS}
    ~~\textbf{Input}: Dataset $(u_k,y_k)$, $k=0,\ldots,N-1$;
    number $N_e\geq 1$ of epochs; initial
    state $x_{0}^s\in\rr^{n_x}$ and 
    covariance matrix $P_{0}^s\in\rr^{n_x\times n_x}$; 
    output noise covariance matrix $R\in\rr^{n_y\times n_y}$
    and process noise covariance matrix $Q\in\rr^{n_x\times n_x}$.
    \vspace*{.1cm}\hrule\vspace*{.1cm}
    \begin{enumerate}[label*=\arabic*., ref=\theenumi{}]
        \item \textbf{For} $e=1,\ldots,N_e$ \textbf{do}:
        \item $x_{0|-1}\leftarrow x_0^s$, $P_{0|-1}\leftarrow P_0^s$;
        \begin{enumerate}[label=\theenumi{}.\arabic*., ref=\theenumi{}.\arabic*]
            \item \textbf{For} $k=0,\ldots,N-1$ \textbf{do}: \hfill\textsf{[forward EKF]}
            \begin{enumerate}[label=\theenumii{}.\arabic*., 
            ref=\theenumii{}.\arabic*]
                \item $C_k\leftarrow \frac{\partial g}{\partial x}(x_{k|k-1},u_k;\theta)$;
                \item $M_k\leftarrow P_{k|k-1}C_k'(R+C_kP_{k|k-1}C_k')^{-1}$;
                \item $e_k\leftarrow y_k-g(x_{k|k-1},u_k;\theta)$;
                \item $x_{k|k}\leftarrow x_{k|k-1}+ M_ke_k$;
                \item $P_{k|k}\leftarrow (I-M_kC_k) P_{k|k-1}(I-M_kC_k)'+M_kRM_k'$;
                \item $A_k\leftarrow \frac{\partial f}{\partial x}(x_{k|k},u_k;\theta)$;
                \item $P_{k+1|k}\leftarrow A_kP_{k|k}A_k'+Q$;
                \item $x_{k+1|k}\leftarrow f(x_{k|k},u_k;\theta)$;
            \end{enumerate}
            \item $x^s_{N}\leftarrow x_{N|N-1}$; $P_N^s\leftarrow P_{N|N-1}$;
            \item \textbf{For} $k=N-1,\ldots,0$ \textbf{do}: \hfill\textsf{[backward RTS smoother]}
            \begin{enumerate}[label=\theenumii{}.\arabic*., 
            ref=\theenumii{}.\arabic*]
                \item $G_k\leftarrow P_{k|k}A_k'P_{k+1|k}^{-1}$;
                \item $x^s_k \leftarrow  x_{k|k} + G_k(x^s_{k+1} - x_{k+1|k})$;
                \item $P^s_k \leftarrow  P_{k|k} + G_k(P^s_{k+1} - P_{k+1|k})G_k'$;
            \end{enumerate}
        \end{enumerate}
        \item \textbf{End}.
    \end{enumerate}
    \vspace*{.1cm}\hrule\vspace*{.1cm}
    ~~\textbf{Output}: Initial state estimate $x^s_0$.
\end{algorithm}


\begin{thebibliography}{10}

\bibitem{AXK17}
B.~Amos, L.~Xu, and J.Z. Kolter.
\newblock Input convex neural networks.
\newblock In {\em Proc. 34th Int. Conf. on Machine Learning. Proceedings of
  Machine Learning Research}, volume~70, pages 146--155, Sydney, Australia,
  2017.

\bibitem{AG07}
G.~Andrew and J.~Gao.
\newblock Scalable training of $\ell_1$-regularized log-linear models.
\newblock In {\em Proc. 24th Int. Conf. on Machine Learning}, pages 33--40,
  2007.

\bibitem{ABO23}
A.Y. Aravkin, R.~Baraldi, and D.~Orban.
\newblock A {Levenberg-Marquardt} method for nonsmooth regularized least
  squares.
\newblock 2023.
\newblock \url{https://arxiv.org/abs/2301.02347}.

\bibitem{AVDBP18}
G.~Armenise, M.~Vaccari, R.~{Bacci Di Capaci}, and G.~Pannocchia.
\newblock An open-source system identification package for multivariable
  processes.
\newblock In {\em UKACC 12th Int. Conf. Control}, pages 152--157, Sheffield,
  UK, 2018.

\bibitem{BTS21}
G.~Beintema, R.~Toth, and M.~Schoukens.
\newblock Nonlinear state-space identification using deep encoder networks.
\newblock In {\em Proc. Machine Learning Research}, volume 144, pages 241--250,
  2021.

\bibitem{Bem23b}
A.~Bemporad.
\newblock Recurrent neural network training with convex loss and regularization
  functions by extended {Kalman} filtering.
\newblock {\em IEEE Transactions on Automatic Control}, 68(9):5661--5668, 2023.

\bibitem{Bem23d}
A.~Bemporad.
\newblock Training recurrent neural-network models on the industrial robot
  dataset under $\ell_1$-regularization.
\newblock In {\em 7th Workshop on Nonlinear System Identification Benchmarks},
  Eindhoven, The Netherlands, April 2023.

\bibitem{Bem23e}
A.~Bemporad.
\newblock Training recurrent neural networks by sequential least squares and
  the alternating direction method of multipliers.
\newblock {\em Automatica}, 156:111183, October 2023.

\bibitem{JAXopt}
M.~Blondel, Q.~Berthet, M.~Cuturi, R.~Frostig, S.~Hoyer, F.~Llinares-L{\'o}pez,
  F.~Pedregosa, and J.-P. Vert.
\newblock Efficient and modular implicit differentiation.
\newblock {\em arXiv preprint arXiv:2105.15183}, 2021.

\bibitem{BBM17}
F.~Borrelli, A.~Bemporad, and M.~Morari.
\newblock {\em Predictive control for linear and hybrid systems}.
\newblock Cambridge University Press, 2017.

\bibitem{JAX}
J.~Bradbury, R.~Frostig, P.~Hawkins, M.J. Johnson, C.~Leary, D.~Maclaurin,
  G.~Necula, A.~Paszke, J.~Vander{P}las, S.~Wanderman-{M}ilne, and Q.~Zhang.
\newblock {JAX}: composable transformations of {P}ython+{N}um{P}y programs,
  2018.

\bibitem{Bro65}
C.G. Broyden.
\newblock A class of methods for solving nonlinear simultaneous equations.
\newblock {\em Mathematics of Computation}, 19(92):577--593, 1965.

\bibitem{BJ68}
R.S. Bucy and P.D. Joseph.
\newblock {\em Filtering for Stochastic Processing with Applications to
  Guidance}.
\newblock Interscience Publishers, New York, 1968.

\bibitem{BCLOS20}
J.V. Burke, F.E. Curtis, A.S. Lewis, M.L. Overton, and L.E.A. Sim{\~o}es.
\newblock Gradient sampling methods for nonsmooth optimization.
\newblock In A.~Bagirov et~al., editor, {\em Numerical Nonsmooth Optimization},
  pages 201--225. 2020.
\newblock \url{http://www.cs.nyu.edu/overton/software/hanso/}.

\bibitem{BLNZ95}
R.H. Byrd, P.~Lu, J.~Nocedal, and C.~Zhu.
\newblock A limited memory algorithm for bound constrained optimization.
\newblock {\em SIAM Journal on Scientific Computing}, 16(5):1190--1208, 1995.

\bibitem{CMO17}
F.E. Curtis, T.~Mitchell, and M.L. Overton.
\newblock A {BFGS-SQP} method for nonsmooth, nonconvex, constrained
  optimization and its evaluation using relative minimization profiles.
\newblock {\em Optimization Methods and Software}, 32(1):148--181, 2017.
\newblock \url{http://www.timmitchell.com/software/GRANSO/}.

\bibitem{KB14}
D.P. Kingma and J.~Ba.
\newblock Adam: A method for stochastic optimization.
\newblock {\em arXiv preprint arXiv:1412.6980}, 2014.

\bibitem{LN89}
D.C. Liu and J.~Nocedal.
\newblock On the limited memory {BFGS} method for large scale optimization.
\newblock {\em Mathematical programming}, 45(1-3):503--528, 1989.

\bibitem{Lju99}
L.~Ljung.
\newblock {\em System Identification : Theory for the User}.
\newblock Prentice Hall, 2 edition, 1999.

\bibitem{Lju01}
L.~Ljung.
\newblock {\em System Identification Toolbox for {MATLAB}}.
\newblock The Mathworks, Inc., 2001.
\newblock \url{https://www.mathworks.com/help/ident}.

\bibitem{LATS20}
L.~Ljung, C.~Andersson, K.~Tiels, and T.B. Sch{\"o}n.
\newblock Deep learning and system identification.
\newblock {\em IFAC-PapersOnLine}, 53(2):1175--1181, 2020.

\bibitem{MB21}
D.~Masti and A.~Bemporad.
\newblock Learning nonlinear state-space models using autoencoders.
\newblock {\em Automatica}, 129:109666, 2021.

\bibitem{MRD18}
D.Q. Mayne, J.B. Rawlings, and M.M. Diehl.
\newblock {\em Model Predictive Control: Theory and Design}.
\newblock Nob Hill Publishing, LCC, Madison,WI, 2 edition, 2018.

\bibitem{VD95}
P.~Van Overschee and B.~De Moor.
\newblock A unifying theorem for three subspace system identification
  algorithms.
\newblock {\em Automatica}, 31(12):1853--1864, 1995.

\bibitem{VD94}
P.~Van Overschee and B.L.R.~De Moor.
\newblock {N4SID}: Subspace algorithms for the identification of combined
  deterministic-stochastic systems.
\newblock {\em Automatica}, 30(1):75--93, 1994.

\bibitem{PGA15}
G.~Pannocchia, M.~Gabiccini, and A.~Artoni.
\newblock Offset-free {MPC} explained: novelties, subtleties, and applications.
\newblock {\em IFAC-PapersOnLine}, 48(23):342--351, 2015.

\bibitem{PAGLRS23}
G.~Pillonetto, A.~Aravkin, D.~Gedon, L.~Ljung, A.H. Ribeiro, and T.B.
  Sch{\"o}n.
\newblock Deep networks for system identification: a survey.
\newblock {\em arXiv preprint 2301.12832}, 2023.

\bibitem{PF94}
G.V. Puskorius and L.A. Feldkamp.
\newblock Neurocontrol of nonlinear dynamical systems with {Kalman} filter
  trained recurrent networks.
\newblock {\em IEEE Transactions on Neural Networks}, 5(2):279--297, 1994.

\bibitem{SSBCV17}
H.~Salehinejad, S.~Sankar, J.~Barfett, E.~Colak, and S.~Valaee.
\newblock Recent advances in recurrent neural networks.
\newblock 2017.
\newblock \url{https://arxiv.org/abs/1801.01078}.

\bibitem{SS23}
S.~S{\"a}rkk{\"a} and L.~Svensson.
\newblock {\em Bayesian filtering and smoothing}, volume~17.
\newblock Cambridge University Press, 2023.

\bibitem{SKS12}
M.~Schmidt, D.~Kim, and S.~Sra.
\newblock Projected {Newton}-type methods in machine learning.
\newblock In S.~Sra, S.~Nowozin, and S.J. Wright, editors, {\em Optimization
  for Machine Learning}, pages 305--329. MIT Press, 2012.

\bibitem{SL19}
J.~Schoukens and L.~Ljung.
\newblock Nonlinear system identification: A user-oriented road map.
\newblock {\em IEEE Control Systems Magazine}, 39(6):28--99, 2019.

\bibitem{SMWN16}
M.~Schoukens, P.~Mattson, T.~Wigren, and J.-P. No{\"e}l.
\newblock Cascaded tanks benchmark combining soft and hard nonlinearities.
\newblock Technical report, Eindhoven University of Technology, 2016.

\bibitem{STP17}
L.~Stella, A.~Themelis, and P.~Patrinos.
\newblock Forward--backward quasi-{Newton} methods for nonsmooth optimization
  problems.
\newblock {\em Computational Optimization and Applications}, 67(3):443--487,
  2017.
\newblock \url{https://github.com/kul-optec/ForBES}.

\bibitem{WGUR22}
J.~Weigand, J.~G{\"o}tz, J.~Ulmen, and M.~Ruskowski.
\newblock Dataset and baseline for an industrial robot identification
  benchmark.
\newblock 2022.
\newblock \url{https://www.nonlinearbenchmark.org/benchmarks/industrial-robot}.

\bibitem{YL06}
M.~Yuan and Y.~Lin.
\newblock Model selection and estimation in regression with grouped variables.
\newblock {\em Journal of the Royal Statistical Society: Series B (Statistical
  Methodology)}, 68(1):49--67, 2006.

\end{thebibliography}
\end{document}